\documentclass[%
 reprint,
 amsmath,amssymb,
 aps,
]{revtex4-2}
\usepackage{graphicx}
\usepackage{dcolumn}
\usepackage{bm}
\usepackage{physics}
\usepackage{mathtools}
\usepackage{float}
\usepackage[usenames,dvipsnames]{color}
\usepackage[colorlinks=false, pdfborder={0 0 1}]{hyperref}
\usepackage{xcolor}

\begin{document}

\preprint{APS/123-QED}

\title{All-optical turbulence mitigation for free-space quantum key distribution using stimulated parametric down-conversion}

\author{A.A. Aguilar-Cardoso}
\email{aagui026@uottawa.ca}
\author{C. Li}%
\author{T.J.B. Luck}
\author{M.F. Ferrer-Garcia}
\author{J. Upham}

\affiliation{
Nexus for Quantum Technologies, Department of Physics, University of Ottawa, Ottawa, K1N 6N5, ON, Canada.
}%

\author{J.S. Lundeen}%
\affiliation{
Nexus for Quantum Technologies, Department of Physics, University of Ottawa, Ottawa, K1N 6N5, ON, Canada.
}%
\affiliation{
 National Research Council of Canada, 100 Sussex Drive, Ottawa, Ontario K1A 0R6, Canada.
}%
\author{R.W. Boyd}
\affiliation{
Nexus for Quantum Technologies, Department of Physics, University of Ottawa, Ottawa, K1N 6N5, ON, Canada.
}%
\affiliation{
Institute of Optics, University of Rochester, Rochester, 14627, NY, USA.
}%

\date{\today}

\begin{abstract}

In this work, we propose and demonstrate a turbulence-resilient scheme for free-space quantum communication. By leveraging the phase conjugation property of stimulated parametric down-conversion, our scheme enables all-optical dynamic correction of spatial-mode distortion induced by atmospheric turbulence, thereby enhancing the secure key rate in high-dimensional quantum key distribution. We develop a theoretical model that provides detailed guidelines for selecting the optimal basis and spatial properties needed to maximize the efficiency of the proposed scheme. Both numerical simulations and experimental results show that, even under strong turbulence, our scheme can reduce the quantum error rates well below the security threshold. These results highlight the potential of nonlinear optical approaches as powerful tools for robust quantum communication in realistic free-space environments. Our work could have important implications for the practical implementation of secure quantum channels over long free-space distances.
\end{abstract}

\maketitle
\section{Introduction}

Quantum key distribution (QKD) enables two distant parties to share encryption keys with information-theoretic security, making it a cornerstone of proposed future communication networks~\cite{bennett-2014}. While QKD was initially proposed on two-level quantum systems, high-dimensional QKD exploits larger Hilbert spaces, providing increased information capacity per photon, enhanced resilience to noise, and improved security against certain classes of attacks~\cite{bechmann-pasquinucci-2000,cerf-2002,bradler-2016}. Different photonic degrees of freedom have been harnessed to implement high-dimensional QKD, with information encoded in frequency, time-bin, or transverse spatial modes~\cite{chang-2025,vagniluca-2020,xu-2020}. In particular, transverse spatial modes, such as Hermite–Gaussian (HG) and Laguerre–Gaussian (LG) modes, retain their well-defined spatial structure during propagation under ideal conditions. This feature allows spatial modes of light to preserve the encoded information over long distances, making them attractive for free-space high-dimensional QKD implementations~\cite{mafu-2013,bouchard-2018}.

However, for QKD implementations in realistic free-space optical (FSO) scenarios, random dynamic fluctuations in the refractive index of air, caused by variations in temperature and pressure, induce phase and amplitude distortions in the propagating spatial modes~\cite{andrews-2005}. These distortions reduce the mode fidelity of the transmitted states, impairing the retrieval of the encoded information at the receiver end~\cite{peters-2025,doster-2016,lochab-2019}.

Several solutions have been proposed and implemented to mitigate the effects of atmospheric turbulence in spatial mode transmission, such as digital adaptive optics and optical phase conjugation techniques. Adaptive optics methods use light modulation devices, such as spatial light modulators (SLM) and deformable mirrors, to apply a digitally computed post-correction mask to the transmitted beam based on the channel distortion characterized by a probe beam. However, they typically require intensive data processing, and their effectiveness is limited by the computational time needed to both analyze and correct the optical distortions before the atmospheric conditions change again. Moreover, the response time of a commercial SLM for phase correction is typically limited to about 100 Hz~\cite{zhou-2021-free}, whereas that of the deformable mirrors is up to 200 Hz~\cite{scarfe-2025,zhao-2020}. In contrast, the characteristic fluctuation rate of dynamic atmospheric turbulence is on the order of 1 kHz for strong turbulence changes~\cite{andrews-2005}. These limitations constrain the effectiveness of the adaptive optics techniques, especially for real-time aberration correction under strong turbulence. Recently, nonlinear optical processes such as four-wave mixing (FWM) have shown promising results due to their capacity of introducing optical phase-conjugation, which conjugates the phase of an input beam and can be used to offset phase distortions acquired in the transmission channel~\cite{zhou-2023,zhou-2025}. However, the response times of many commonly used $\chi^{(3)}$ materials for FWM have not been sufficiently fast, with measured values on the order of 10 mHz. Although novel $\chi^{(3)}$ materials promise ultrafast FWM responses of 1 THz, characterizing their optical properties and achieving practical nonlinear conversion efficiency with them remain works in progress~\cite{alam-2016}. 

Recently, stimulated parametric down-conversion (StimPDC) has gained attention owing to its inherent phase-conjugation property. Importantly, studies have demonstrated that StimPDC can generate phase-conjugation of partially coherent beams, making this technique particularly suitable for correcting beams that have been distorted by turbulent media~\cite{santos-2021,santos-2025}. Furthermore, this three-wave mixing nonlinear process has been shown to correct spatial modes affected by specific optical aberrations~\cite{singh-2024}, as well as to recover images distorted by turbulence~\cite{zou-2014}. Although StimPDC-based optical phase conjugation is considered more experimentally challenging than its FWM-based counterpart~\cite{yariv-1978}, StimPDC offers an advantage for correcting dynamic aberrations due to its faster nonlinear response time. In particular, the nonlinear response time of $\chi^{(2)}$ materials has been demonstrated to be ultrafast, exceeding 1 THz, in commercially available nonlinear crystals~\cite{wang-2025}.  

\begin{figure*}
\centering
\includegraphics[width=1\linewidth]{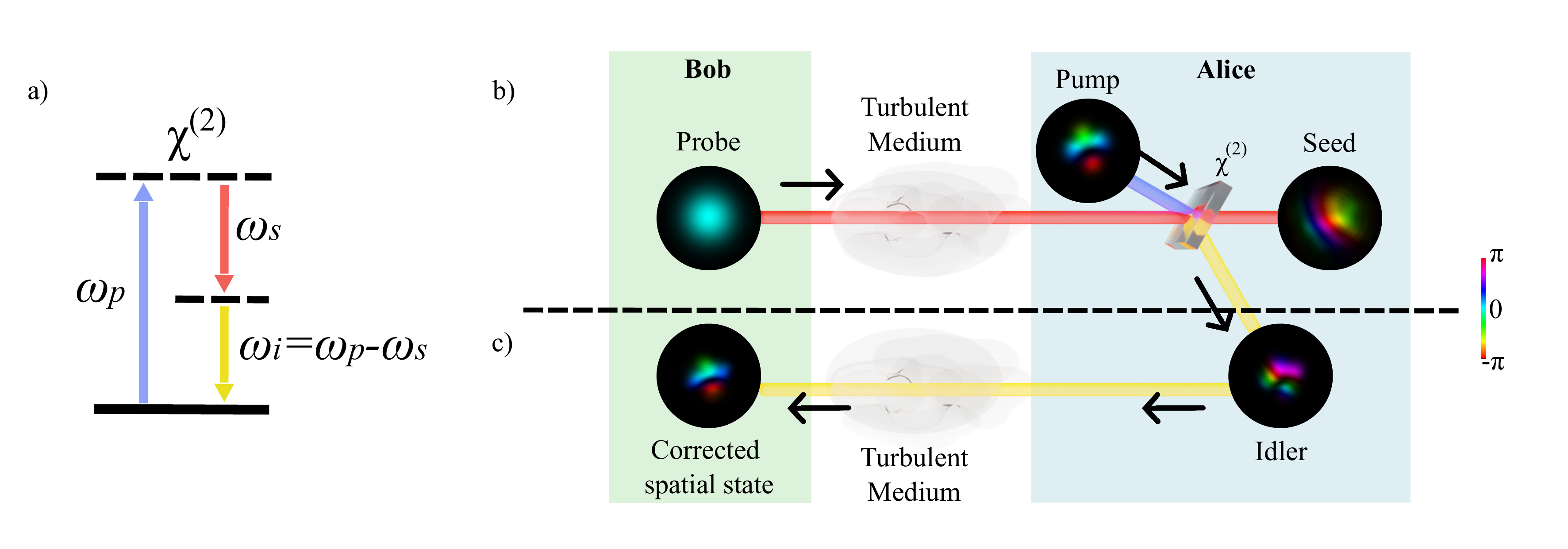}
    \caption{a) Energy level description of StimPDC. b) StimPDC scheme: Bob sends a Gaussian beam through a turbulent transmission channel to probe the turbulence, and thereby acquires phase distortions. Alice pumps a thin nonlinear crystal with a laser beam encoded with the spatial mode she wants to transmit to Bob. She also seeds the crystal along the signal path with the distorted probe beam. c) As a result, the idler beam carries Alice’s target transmission mode together with the phase-conjugate of the turbulence distortions. For the degenerate situation in which the probe and idler beams have the same wavelength, phase distortions of the returning beam are cancelled, and the intended spatial state is received.}
    \label{fig:aliceybob}
\end{figure*}

In this work, we propose and demonstrate the first all-optical self-correction scheme for turbulence-resilient free-space high-dimensional QKD using StimPDC. Specifically, by exploiting the intrinsic phase conjugation property of the StimPDC (Fig.~\ref{fig:aliceybob}), we combine the spatial structure of the quantum state intended for transmission with the conjugate phase information of the turbulent channel, thereby embedding the correction within the nonlinear interaction. Unlike previous approaches that involve channel characterization in advance, our scheme does not require either party to have prior knowledge of the channel, as the nonlinear process inherently performs the compensation. We test our scheme’s performance under varying turbulence conditions through numerical simulations and demonstrate its feasibility and effectiveness using a proof-of-principle experiment. These results offer a promising foundation for developing turbulence-resilient quantum FSO systems and could have implications for long-distance QKD and remote quantum networks.

\section{Principles of the StimPDC scheme}
\label{protocol}

Typically, the prepare-and-measure (P\&M) QKD scheme for spatial modes of light (eigenmodes of the paraxial wave equation) relies on Alice’s ability to efficiently transmit to Bob several sets of states $\ket{U^{(b)}}=\{ \ket{u^{(b)}_1},\ket{u^{(b)}_2},...,\ket{u^{(b)}_d} \}$ that each forms an orthonormal basis $b$ of dimension $d$; i.e.,
\begin{equation}
\label{ortoQuantum}
\braket{u^{(n)}_{j}|u^{(n)}_{j'}}=\delta_{j,j'}, \ \ \ j,j'=1,...,d, \ \ \ n=1,...,b,
\end{equation}
and must be mutually unbiased bases (MUBs) with one another, meaning that the states that form each basis satisfy~\cite{mafu-2013,bouchard-2018}:
\begin{equation}
\label{MUBcondiiton}
    |\braket{u^{(n)}_{j}|u^{(n')}_{j'}}|^2=\frac{1}{d}, \ \ \ j,j'=1,...,d, \ \ \ n\neq n'=1,...,b.
\end{equation}

This property ensures that measurements in a mismatched basis yield completely random outcomes. As a result, if an eavesdropper attempts to extract information using a basis different from the one used for transmission, they will be unable to gain any meaningful information. In this work, without loss of generality, we focus on the BB84-type protocols, in which one uses two sets of MUBs, each of dimension $d$.

In FSO scenarios, the phase aberrations introduced by atmospheric turbulence can be modeled by a spatially varying phase function $\phi_T(\Vec{r},\omega)$, which modifies and distorts the spatial profile of the state as a function of transverse position $\Vec{r}$ and frequency $\omega$~\cite{khare-2020,andrews-2005}. A key measure of the system’s performance is the crosstalk matrix $M^b_{j,j'} = |\braket{u^{(b)}_{j,\text{Bob}}|u^{(b)}_{j',\text{Alice}}}|^2$, which quantifies the overlap between the transmitted state $\ket{u^{(b)}_{j',\text{Alice}}}$ and the detected state $\ket{u^{(b)}_{j,\text{Bob}}}$. From this matrix, one can calculate the quantum error rate (QER) of the basis $b$ induced by the turbulence, as the fraction of erroneous detections relative to the total number of detection events in that basis:
\begin{equation}
\label{Qerror}
    Q_b = 1 - \frac{1}{d} \sum_{j=1}^d M^b_{j,j}.
\end{equation}

The average QER across all bases yields the total QER of the communication, denoted as $Q$. This metric reflects the average deviation from perfect orthogonality in the preparation and detection of spatial states, thereby serving as a global figure of merit for the communication link between Alice and Bob. 

From the total QER value, the secure key rate $r$ can be calculated using the expression~\cite{bradler-2016}:

\begin{equation}
\label{rate}
    r=\text{log}_2(d)+2(1-Q)\text{log}_2(1-Q)+2Q\text{log}_2\left(\frac{Q}{d-1}\right).
\end{equation}

The maximum tolerable error is determined by the value $Q_{max}$ at which the secure key rate $r$ drops to zero. This threshold increases with the dimension $d$, making high-dimensional protocols more resilient to communication errors and noise, thereby enhancing their practical robustness in real-world quantum communication scenarios~\cite{liao-2017,meyer-2025,shi-2025,mirhosseini-2015}.

We propose a scheme based on StimPDC to overcome the undesired phase distortions introduced by atmospheric turbulence on the transmitted spatial states (i.e. to reduce $Q$). StimPDC is a nonlinear optical process in which a pump beam with frequency $\omega_p$, interacts with a second-order nonlinear medium with a nonlinear susceptibility of $\chi^{(2)}$ to generate a signal and an idler photon. The process is seeded with a signal beam with frequency $\omega_s$, which enhances the generation of idler photons at frequency $\omega_i=\omega_p-\omega_s$ (Fig.~\ref{fig:aliceybob}a) ~\cite{ribeiro-1999}. Crucially, the idler beam is related to the phase-conjugate of the seed beam. The scheme we propose is depicted in Fig.~\ref{fig:aliceybob}b-c, and is implemented as follows:

In contrast to the P\&M scheme, Alice does not directly send her prepared states. Instead, she pumps a nonlinear crystal $\chi^{(2)}$ with a coherent state $\ket{U_A}$, prepared in a spatial mode $U_A(\vec{r})$ that belongs to a family of spatial modes satisfying the conditions given by Eqs. \ref{ortoQuantum} and \ref{MUBcondiiton}. This beam is characterized by beam waist $w_A$, Rayleigh range $z_{RA}$, and frequency $\omega_A$. Meanwhile, Bob independently sends a coherent state $\ket{U_B}$, prepared in a Gaussian mode $U_B(\vec{r})$ with waist parameter $w_B$, Rayleigh range $z_{RB}$ and frequency $\omega_B$ through the shared turbulence channel of length $Z_T$ connecting him with Alice. As it propagates toward Alice’s stage, Bob’s beam undergoes wavefront distortions, now denoted as $\ket{U^T_B}$, thereby serving as a probe of the turbulence effects within the channel. Upon receiving the distorted probe beam from Bob, Alice seeds it into the signal path of the nonlinear process to stimulate the idler photon generation with frequency $\omega_i=\omega_A-\omega_B$ (Fig.~\ref{fig:aliceybob}b). 

Assuming both the pump and seed are in the low-gain regime, the unnormalized stimulated contribution to the idler state obtained by Alice is~\cite{de-oliveira-2020}:

\begin{equation}
\label{stateStim}
    \ket{U_{i}} \propto \iint d\Vec{q_s}d\Vec{q_i} \Tilde{U}_A(\Vec{q_s},\Vec{q_i})\Tilde{U}_B^{T*}(\Vec{q}_s)\ket{1,\Vec{q_i}},
\end{equation}
in which we have written the state vector in Fock state representation (photon number = 1), with $\Vec{q_{s(i)}}$ representing the transverse momentum of the signal(idler) beam, while $\Tilde{U}_A(\Vec{q})$ and $\Tilde{U}_B(\Vec{q})$ represent the angular spectrum of the pump and seed beam, respectively. As seen in Eq.~\ref{stateStim}, the spatial structure of the idler state is determined by the product of the spatial profile Alice aims to transmit and the conjugated phase information acquired by the probe beam as it propagates through the turbulent medium. It is worth noting that Eq.~\ref{stateStim} describes the stimulated contribution to the idler field under the low-gain approximation, where only single-photon output states are considered and higher-order multi-photon terms are neglected due to their negligible contribution to the overall field~\cite{de-oliveira-2020}. Even though the StimPDC process can generate multi-photon states, the same QKD principles remain valid when using weak coherent beams, which are commonly employed in real QKD implementations~\cite{liao-2017,meyer-2025,shi-2025}.

In the final step of the scheme, Alice then sends the idler beam back to Bob (Fig.~\ref{fig:aliceybob}c). In the degenerate case, this idler beam has the same frequency as Bob's beam ($\omega_i=\omega_B$). This requirement is essential, since the phase aberrations introduced by the turbulent medium are considered dispersive, that is, inherently frequency-dependent, $\phi_T(\Vec{r}, \omega)$. This conjugated part of the idler beam is able to mitigate arbitrary phase distortions acquired during the return path, effectively undoing the undesired phase term initially accumulated. 

\section{Engineering spatial states with StimPDC}
\label{engene}

The proposed scheme requires further optimization to fully exploit the distortion-correction capabilities of StimPDC. In particular, the choice of spatial basis for information transmission and the beam parameters of the seed, pump, and idler beams must be matched to the characteristics of the turbulent channel.

The reliability of spatial mode transfer via StimPDC, in the absence of turbulence, is quantified by the fidelity between the generated idler beam state $\ket{U_i}$ and the pumped target state $\ket{U_A}$ to be transmitted by Alice. This fidelity is given by:
\begin{equation}
\label{fidelity}
    F = |\braket{U_A|U_i}|^2=\frac{\left|\int U_A^*(\vec{r}) U_B^*(\vec{r}) U_A(\vec{r}) d\vec{r}\right|^2}{\int |U_A(\vec{r})|^2 d\vec{r} \int |U_B^*(\vec{r}) U_A(\vec{r})|^2 d\vec{r}},
\end{equation}
where normalization factors have been included for each state. As demonstrated in Ref.~\cite{aguilar}, the fidelity of spatial mode transfer depends solely on the ratio of the beam waists $\gamma = w_B / w_A$. High fidelities require $\gamma > 1$, meaning the seed beam must have a larger transverse extent than the target mode. This condition makes the overall fidelity of mode transmission dependent on the choice of basis. For instance, LG modes are less ideal choices since the sizes of their transverse profiles increase with mode orders, resulting in diminishing transmission fidelity at higher mode orders~\cite{aguilar}. However, one can overcome this issue by using mode families in which all elements share the same optical diameter $D$, such as balanced superpositions of LG modes. We define $D$ as the second-moment diameter of the spatial mode that will be transmitted through the channel~\cite{cocotos-2025}.

\begin{figure}
\centering
    \includegraphics[width=1\linewidth]{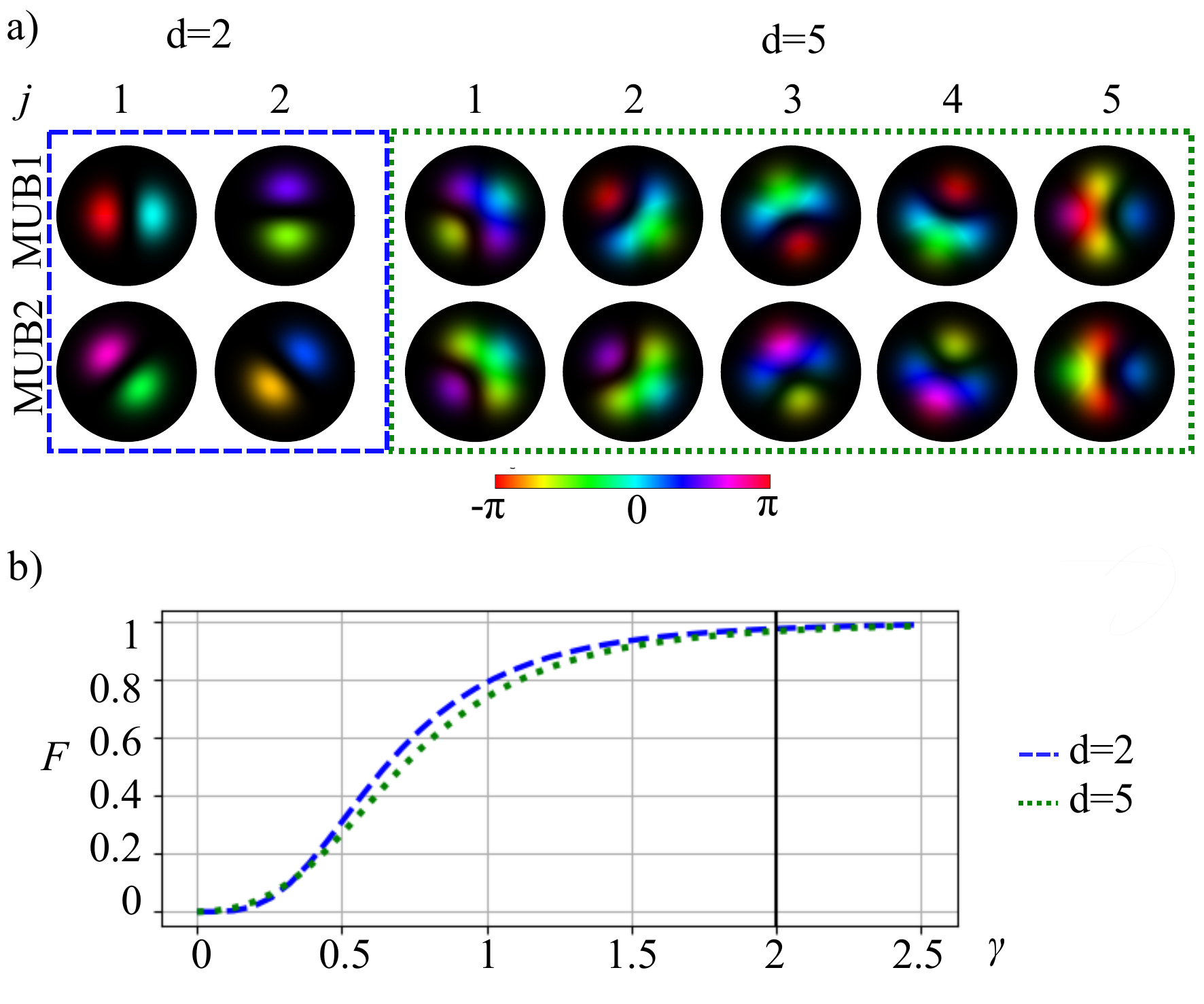}
    \caption{a) Amplitude and phase structure of the designed MUBs spatial states generated via StimPDC for $d=2$ and $d=5$. b) Fidelity of the spatial basis generated with StimPDC as a function of the ratio $\gamma = w_B / w_A$. We see that for $\gamma=2$ we obtain high fidelities for every element of each basis.}
    \label{fig:modes}
\end{figure}

We test our scheme using two sets of MUBs, namely MUB1 and MUB2. We use $d=2$ to evaluate our scheme in a typical two-dimensional QKD system, and $d=5$ to evaluate its performance in a high-dimensional QKD configuration. Therefore, we define MUB1 and MUB2 for each dimension. The states forming both MUBs are defined as superpositions of LG modes with zero radial index $q=0$, with $l \in [-1, 1]$ for $d=2$ and $l \in [-2, 2]$ for $d=5$. The procedure for constructing both bases in each dimension is detailed in the Supplementary Material. These states are shown in Fig.~\ref{fig:modes}a. These bases satisfy both the orthonormality condition (Eq.~\ref{ortoQuantum}) and the mutual unbiasedness condition (Eq.~\ref{MUBcondiiton}). The calculation of the fidelity, given by Eq.~\ref{fidelity}, for these states is shown in Fig.~\ref{fig:modes}b, where we find identical fidelity behavior across all elements of each basis. For $\gamma = 1$, we obtain $F = 0.79$ for $d=2$ and $F = 0.73$ for $d=5$, increasing to $F = 0.97$ for both dimensions at $\gamma = 2$. 

We therefore choose $\gamma = 2$ as the value for generating the desired spatial states with StimPDC. Under this condition, and with $\omega_B = \omega_i$, the remaining adjustable parameters are the beam waists $w_B$ and $w_i$, which determine the optical diameters $D_B$ and $D_i$. The optical diameter of a beam with OAM is given by~\cite{cocotos-2025}:

\begin{equation}
\label{diameter}
    D(z)=2w_0\sqrt{\left(|l|+1\right)\left(1+\left(\frac{z}{z_R}\right)^2\right)},
\end{equation}
where $l$ is the topological charge and $z_R=\frac{\pi w_0^2}{\lambda}$ the Rayleigh range. In our case, the probe is a Gaussian beam ($l=0$), while the transmitted spatial modes are superpositions of OAM states with maximum $|l|=2$.

\begin{figure}[h]
\centering
    \includegraphics[width=1\linewidth]{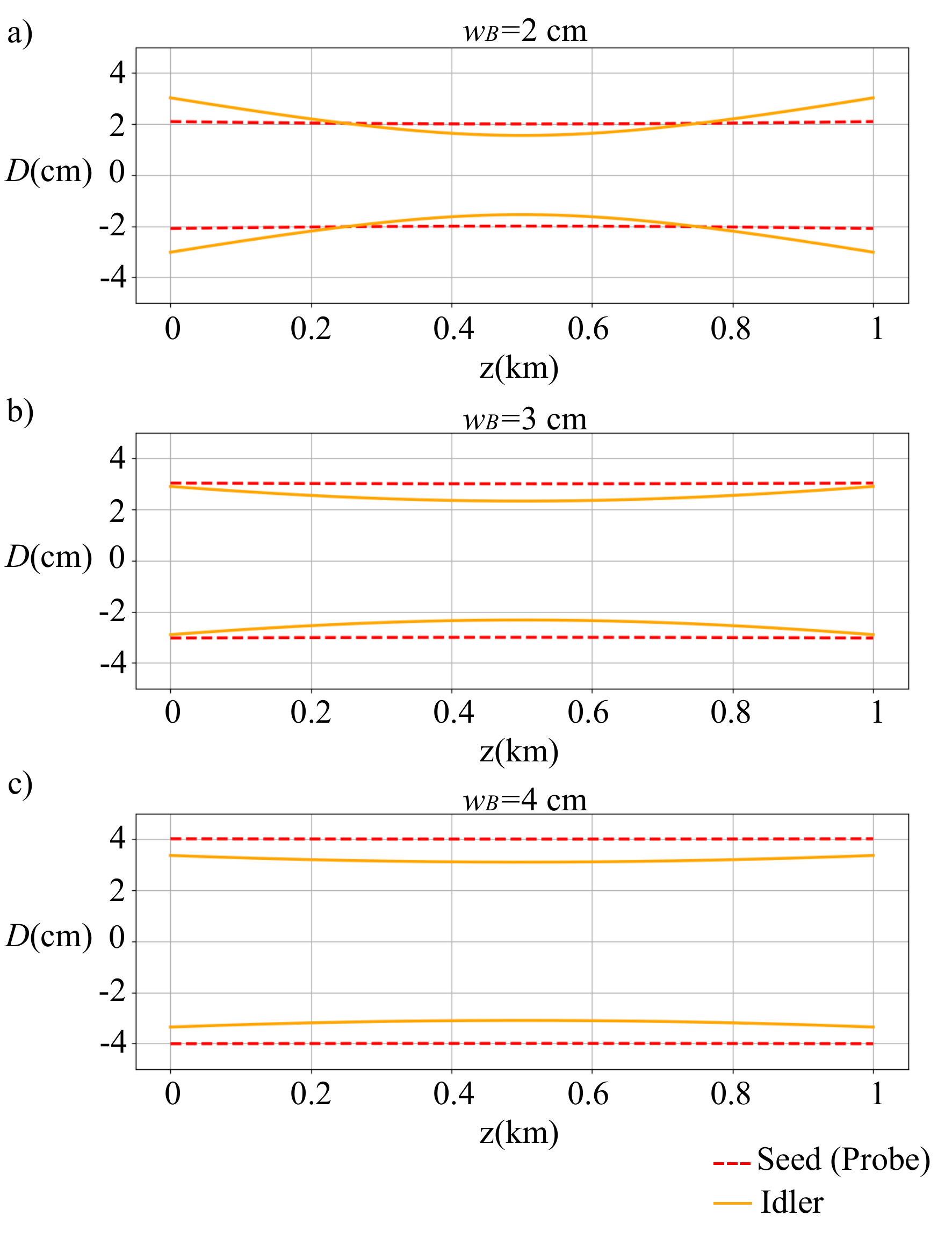}
    \caption{ Optical diameter of the probe beam sent by Bob and the stimulated idler beam generated by Alice, shown for several different probe beam waists: a) 2 cm, b) 3 cm, and c) 4 cm. There exists a specific value of $w_B$ for which the condition $D_B(Z_T) = D_i(Z_T)$ is satisfied, corresponding in this case to scenario b). Although scenario c) satisfies the proposed criterion to a greater extent, it also amplifies the aberrations introduced by turbulence.}
    \label{fig:divergence}
\end{figure}

In the scheme, Bob has control over the probe beam’s waist $w_B$ and its corresponding Rayleigh range $z_B$. However, the idler beam generated via StimPDC possesses a different waist $w_i$ and, consequently, a different Rayleigh range $z_i$. These quantities can be evaluated within a classical model of StimPDC, as shown in Ref.~\cite{aguilar}, since the detection probability of $\ket{U_{i}}$ corresponds to the  idler beam intensity in the classical model, given by $|U_A(\vec{r})U^*_B(\vec{r})|^2$~\cite{de-oliveira-2020}. Consequently, we can compute the beam diameters $D_B(z)$ and $D_i(z)$ at an arbitrary distance $z$ along the propagation path. Fig.~\ref{fig:divergence} shows the analytically computed beam diameters of both the probe and the idler beams for different $w_B$ values, for $\gamma=2$ and $\lambda_B=810$ nm, for $Z_T =$ 1 km. 

For the probe beam to adequately sample the spatial extent of the idler beam along the turbulent path, we require $D_B(z) \geq D_i(z)$ for $z\in[0,Z_T]$. However, one cannot arbitrarily enlarge $w_B$ (i.e. $D_B$) as doing so would exacerbate the effects of turbulence. The reason behind this constraint is that the strength of turbulence experienced by an optical beam depends on the ratio between the beam's optical diameter $D$ and the Fried parameter $r_0= \left( 0.423 C_n^2 k^2 Z_T \right)^{-3/5}$, where $C_n^2$ is the refractive index structure constant and $k$ is the optical wavenumber~\cite{Fried-1965, khare-2020,andrews-2005}. One can interpret the quantity $D/r_0$ as the number of atmospheric eddies (given that $r_0$ characterizes the spatial coherence length of the atmospheric eddies) within the beam's transverse profile. Therefore, as $D_B$ increases, so does the strength of turbulence. We note that this condition is not unique to our scheme, as any adaptive optics scheme relying on a beacon beam is subject to the same constraint~\cite{scarfe-2025,zhao-2020}. To overcome this, we adopt a practical optimization criterion: choose $w_B$ such that $D_B(Z_T) = D_i(Z_T)$. As an example, for $Z_T = 1~\mathrm{km}$, this yields $w_B \approx 3~\mathrm{cm}$, as seen in Fig. \ref{fig:divergence}b.

In our scheme, the idler beam has a smaller diameter than the probe beam and thus experiences a different level of turbulence-induced distortions. However, both beams are subject to the same spatial phase modulations $\phi_T(\Vec{r},\omega)$. It is also worth mentioning that it is not possible for Alice to arbitrarily adjust the beam waist and Rayleigh range of the idler beam beam waist, since this would simultaneously modify its transverse phase structure $\phi_T(\Vec{r},\omega)$, which would introduce additional aberrations into the idler beam.

\section{Testing the StimPDC scheme}
\label{tests}
\subsection{Numerical simulations}

\begin{figure*}
\centering
    \includegraphics[width=1\linewidth]{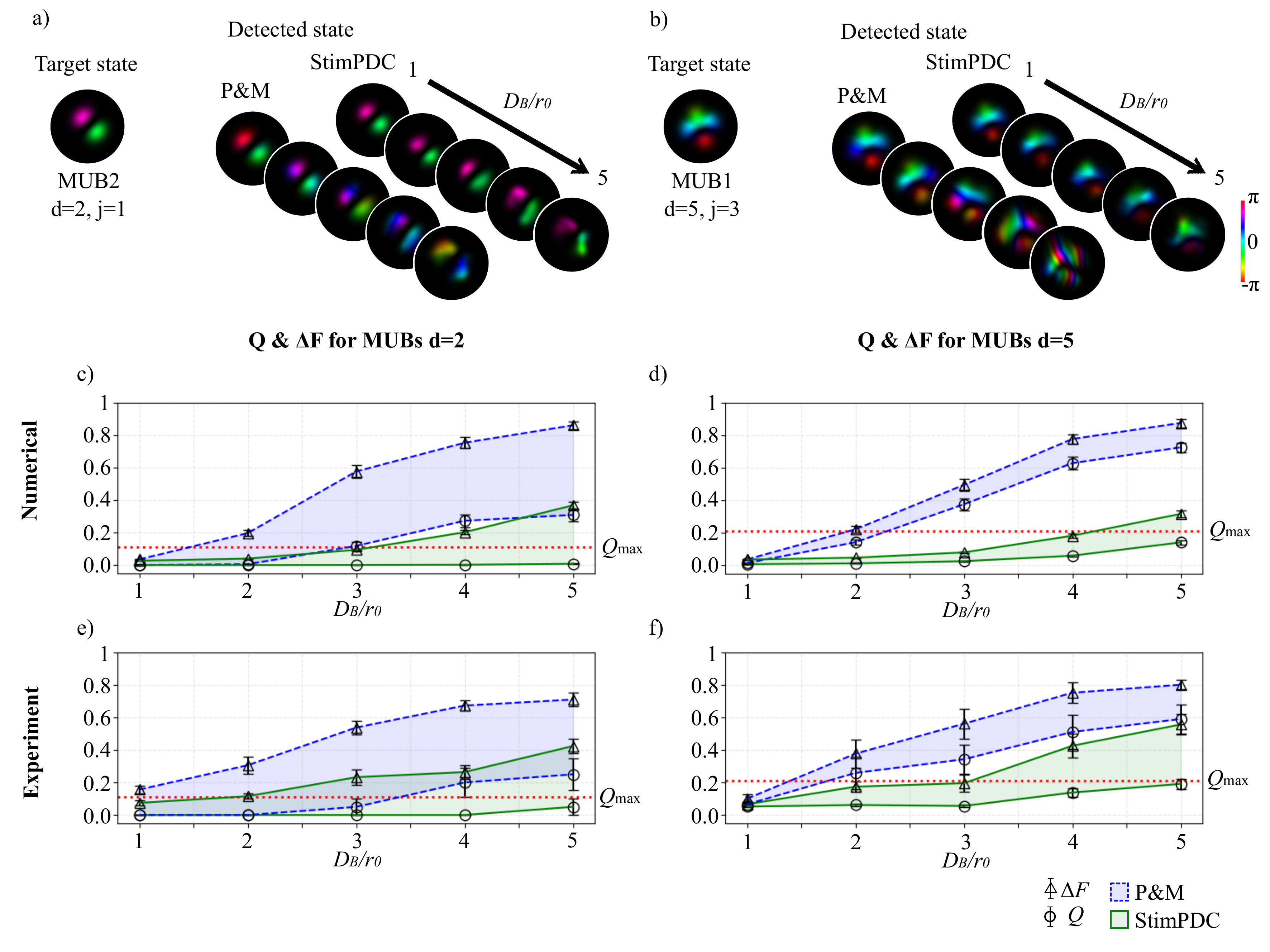}
    \caption{Field simulation results for a) MUB2, $d=2$, $j=1$ and b) MUB1, $d=5$, $j=3$. The target state is shown along with the spatial states distorted by turbulence, and corrected by our scheme. With the StimPDC scheme, it can be observed that the amplitude retains the some aberration as in the P\&M case. However, the phase resembles that of the original beam. Panels c) and d) show QER ($Q$) and fidelity loss ($\Delta F$) for the P\&M and StimPDC scheme using MUB1 and MUB2 for $d=2$ and $d=5$ obtained with the split-step method. Panels e) and f) present the experimental results. The error bars correspond to the standard error. The shaded area highlights the difference between the QER and the fidelity loss, indicating that the latter does not fully account for the observed crosstalk reduction.}
    \label{fig:resultados}
\end{figure*}

We employ the split-step method, a standard numerical method for simulating beam propagation through turbulence~\cite{khare-2020,peters-2025}. Here, the propagation path is divided into small segments where free-space propagation and turbulent phase masks are applied sequentially: each step consists of free-space propagation via the angular spectrum method, followed by a statistically accurate phase screen based on the Kolmogorov power spectrum, representing the spatially varying phase distortions induced by atmospheric turbulence. Repetitively performing these operations models the cumulative effect of turbulence along the entire path, thereby generating both phase and amplitude distortions in the originally propagated beam. Comprehensive details of these simulations, together with the procedures for generating the employed phase masks, are presented in the Supplementary Material.

We test the scheme using the MUBs derived in the previous section, along with their established optimal beam parameters ($\gamma=$ 2, $\lambda_B=$ 810 nm, and $w_B=$ 3 cm for $Z_T=$ 1 km). As a representative example, in Figs. ~\ref{fig:resultados}a and ~\ref{fig:resultados}b, we present simulated amplitude and phase profiles for the case $d=2$ with the state $\ket{u^{(2)}_1}$, and for $d=5$ with the state $\ket{u^{(1)}_3}$. We investigate both cases under turbulence conditions of $D_B/r_0 = [1,5]$, which effectively cover regimes ranging from moderate ($D_B/r_0 = 1$) to strong turbulence ($D_B/r_0 = 5$). We note that our qualitative description of turbulence strength ("moderate/strong") is consistent with that used in existing literature \cite{zhao-2020, Krenn-2016}. Specifically, in Ref.~\cite{zhao-2020}, the turbulence strength of a free-space channel across a university campus with $D_B/r_0 = 1.03$ to $2.98$ is described as "moderate to strong"; and in Ref,~\cite{Krenn-2016}, the turbulence strength of a 143-km channel between two islands, which is characterized by $D=35$~mm and $r_0=1.3$~cm ($D_B/r_0 \approx 2.7$), is referred to as "strong".

To demonstrate the advantage of the StimPDC scheme over the standard P\&M approach, we simulate the transmission of spatial states for both schemes under the same conditions. We then compute the fidelity matrix $F^b_{j,j'}$ as the overlap between the retrieved mode $j$ and the ideal mode $j'$, for all states across both MUBs in each dimension. We evaluate the average fidelity $F$ of both MUBs for $j=j'$. To obtain statistically consistent results, we averaged over 50 realizations of spatial state transmission through turbulence. To better compare the performance of both schemes, we use the average fidelity loss $\Delta F=1-F$ to quantify their resistance to turbulence. For the P\&M approach, we identify that for $D_B/r_0=1$, the fidelity loss remains below 5\%, while for $D_B/r_0=5$, the fidelity loss is nearly complete, increasing to almost 95\%. Figs.~\ref{fig:resultados}c and ~\ref{fig:resultados}d show $\Delta F$ for MUBs with $d=2$ and $d=5$, respectively. For both dimensions, the P\&M method exhibits up to a 90\% fidelity loss, whereas our StimPDC scheme retains fidelity with losses limited to only about 30\%. 

Our scheme demonstrates improved fidelity over the P\&M scheme while retaining some fidelity loss. This is because the turbulence effects are only partially corrected, as the probe beam does not experience the same propagation distortions as the spatial mode under test. However, this fidelity loss is not entirely related to modal crosstalk, but rather from residual aberrations that distort the ideal spatial profile, as seen in the examples of Fig.\ref{fig:resultados}a and \ref{fig:resultados}b. To confirm this, we compute the QER from the normalized crosstalk matrix, defined as $M^b_{j,j'} = F^b_{j,j'}/\sum_{j'} F^b_{j,j'}$~\cite{zhou-2021}.

Figs.~\ref{fig:resultados}c and ~\ref{fig:resultados}d also display the average QER derived from the normalized crosstalk matrices for $d=2$ and $d=5$, respectively. States with $d=2$ show greater resilience to crosstalk, which can be attributed to their HG mode structure. HG modes have a Cartesian spatial profile without a central phase singularity, making them less sensitive to small angular distortions and to the loss of azimuthal symmetry~\cite{gu-2020,restuccia-2016,cox-2019}. Nevertheless, for both dimensions, when $D_B/r_0 > 2$ under the P\&M scheme, crosstalk-induced QER surpasses the maximum tolerable error threshold $Q_{\text{max}}$, rendering information transmission entirely ineffective. In contrast, with our StimPDC scheme, the QER in all cases remains below the critical threshold $Q_{\text{max}}$. 

\begin{figure}[ht]
\centering
    \includegraphics[width=1\linewidth]{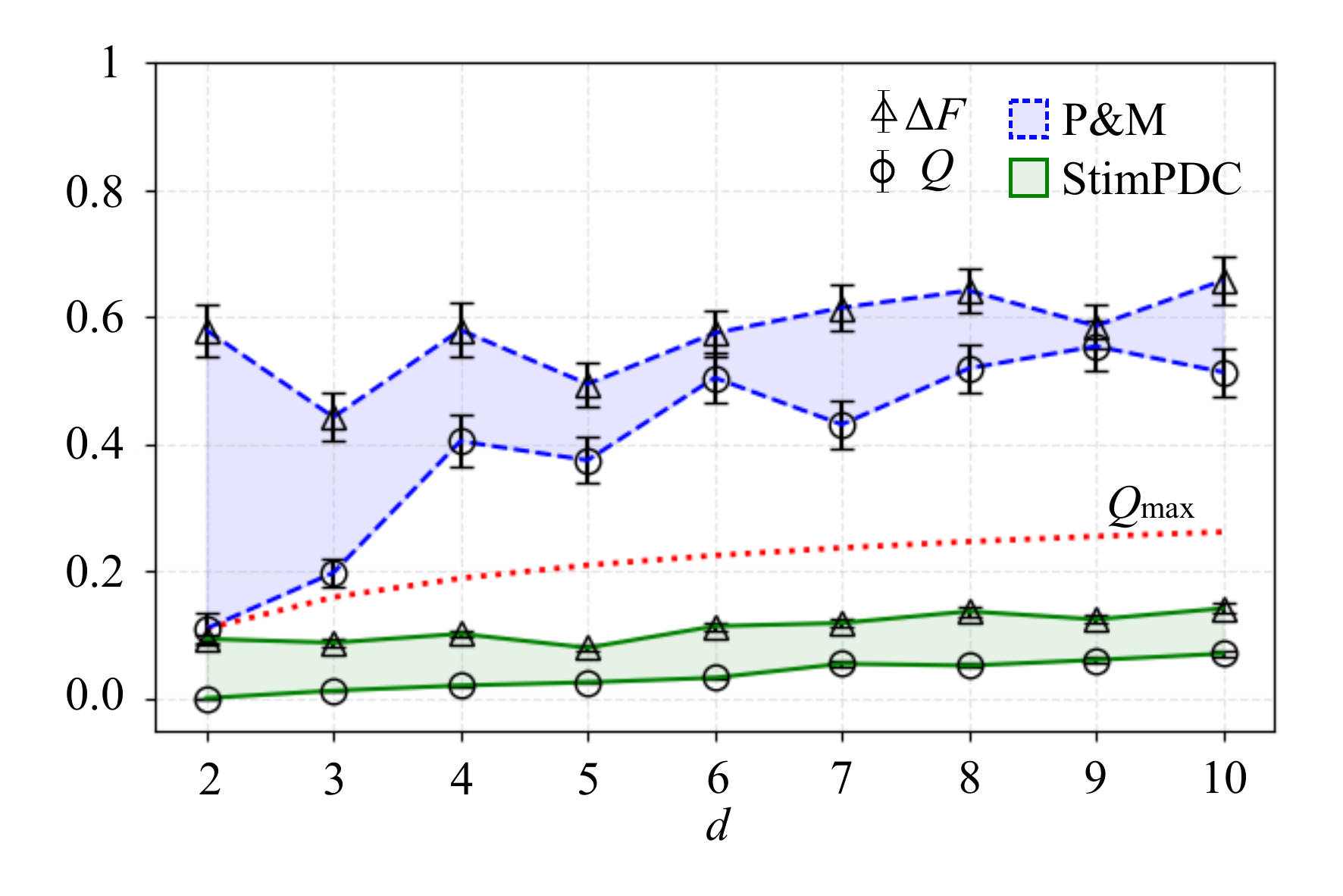}
    \caption{ QER ($Q$) and fidelity loss ($\Delta F$) for the P\&M and StimPDC schemes using MUB1 and MUB2 for $d = [2,10]$, obtained with the split-step method for $D_B/r_0 = 3$. For all dimensions, the StimPDC scheme maintains $Q < Q_{\text{max}}$.}
    \label{fig:dimensionsQ}
\end{figure}

Finally, to demonstrate that the efficiency of our scheme is not dimension-dependent, Fig.~\ref{fig:dimensionsQ} presents the QER and fidelity loss for high-dimensional QKD ($d = [2,10]$) at $D_B/r_0 = 3$, evaluated for different MUB1 and MUB2 sets, obtained with our simulations. We show that, across all dimensions, the QER consistently remains below $Q_{\text{max}}$, confirming the robustness and effectiveness of our protocol in mitigating crosstalk for high-dimensional QKD systems based on spatial states. It is worth noting that, as the dimensionality increases, the used spatial modes become more complex in their phase and amplitude distributions, which causes $\Delta F$ to increase with both turbulence strength and dimension. While it is still important at any dimension to select a basis that is more resilient to turbulence, both the turbulence resilient bases and more susceptible bases will benefit from the StimPDC scheme.

\subsection{Experiment}

\begin{figure*}
\centering
    \includegraphics[width=1\linewidth]{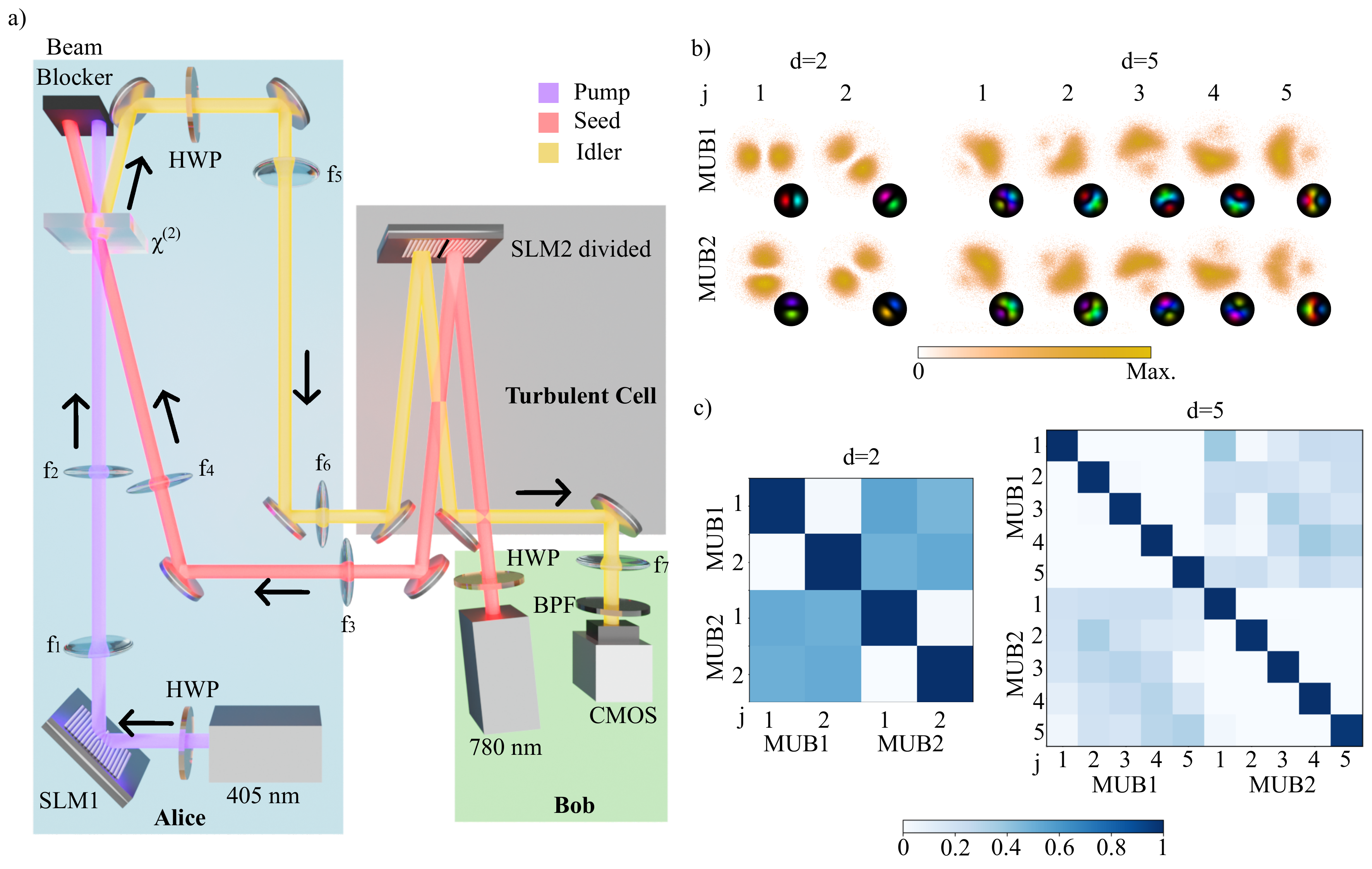}
    \caption{a) Experimental setup. Alice pumps the spatial state she intends to share with Bob with a 405 nm laser, while simultaneously seeds the 780 nm beam sent by Bob through the turbulent cell. This process generates StimPDC at 840 nm. The resulting beam is then sent back through the turbulent cell so that Bob can receive the state. b) Normalized intensities of the idler beam, obtained at the SLM2 plane using the CMOS camera are shown for the previously chosen MUBs in cases of $d=2$ and $d=5$. c) The corresponding normalized crosstalk matrices for both dimensions are also presented, experimentally confirming that the QKD conditions given by Eqs. \ref{ortoQuantum} and \ref{MUBcondiiton} are satisfied.}
    \label{fig:EXPERI}
\end{figure*}

As an experimental proof of principle, we use the setup shown in Fig.~\ref{fig:EXPERI}a. A collimated 405 nm beam illuminates spatial light modulator 1 (SLM1). SLM1 has been calibrated for this wavelength, and is used to encode the desired spatial mode through standard holographic techniques. All selected parameters follow the optimization criteria described in Section~\ref{engene}. The structured pump beam, with power $\approx$ 1 mW, and a beam waist radius $w_A=0.13$ cm ($D_A =0.26\sqrt{3}= 0.45$ cm), is then imaged onto a 2-mm-long beta-barium borate (BBO) crystal (cut for type-II phase matching) using a telescope composed of lenses with focal lengths $f_1=500$ mm and $f_2=250$ mm. A second collimated Gaussian beam at 780 nm, with power $\approx$ 20 mW, and a beam waist radius of $w_B = 0.25$ cm ($D_B = 0.5$ cm), propagates through a free-space path of length $Z_T = 1$ m. At the midpoint, it illuminates the first region of SLM2, where we encode phase masks that simulate atmospheric turbulence if needed. These phase masks are generated using the same numerical method described in the simulation section of the supplementary material, allowing us to recreate the effect of a turbulent channel on the probe beam for $D_A/r_0\in[1,5]$. The beam is then demagnified using a telescope with lenses $f_3=500$ mm and $f_4=250$ mm, and subsequently seeded into the BBO crystal at an angle of 4$^{\circ}$ with respect to the pump beam, resulting in the generation of StimPDC with $\gamma= w_B / w_A\approx 2$.

First, to test the spatial states in the absence of turbulence, we modulate the pump beam with spatial modes within the chosen MUBs using SLM1, generating the MUB spatial states for both $d=2$ and $d=5$. The first region of SLM2 displays a flat, turbulence-free, phase profile. The generated idler beam is collected using a magnifying telescope with lenses $f_5=250$ mm and $f_6=500$ mm, and directed to a second region on SLM2, that is separated from the first region illuminated by the probe beam. Fig.~\ref{fig:EXPERI}b shows the intensity distributions of the generated modes at the plane of SLM2. In this second region of the SLM2, we encode a projection onto the target spatial mode. It is worth noting that since our experiment is carried out in the near-degenerate StimPDC regime, we have calibrated SLM2 to operate at 810 nm, which is the wavelength at degenerate phase-matching. Finally, the beam is sent through a Fourier lens $f_7=500$ mm, and the intensity at the center of the optical axis is measured using a CMOS camera in the Fourier plane. This central intensity provides an estimate of the modal overlap $\left| \braket{u_j^{(b)}|u_{j'}^{(b')}} \right|^2$ \cite{pinnell-2020,meyer-2025}. Please refer to the Supplementary Material for more details. Then, we measure the fidelity and crosstalk matrices in the absence of turbulence. The results are shown in Fig.~\ref{fig:EXPERI}c, experimentally confirming that the QKD conditions given by Eqs.~\ref{ortoQuantum} and \ref{MUBcondiiton} are satisfied.

To evaluate the P\&M scheme under turbulence, we pump the crystal with the selected MUB states using SLM1. The first region of SLM2 still displays a flat, turbulence-free phase profile, allowing us to seed the StimPDC process with an unaberrated Gaussian beam. Meanwhile, turbulent phase masks simulating different turbulence strengths are applied to the second region of SLM2, along with the projection of the target spatial mode. 

To evaluate the StimPDC scheme, we follow the same procedure; however, in this case, we apply the same turbulent phase mask to both the seed and idler beams (i.e., to both regions of SLM2), so that the seed beam now acts as a probe for the turbulent channel. The phase-conjugation then automatically compensates for the influence of turbulence on the idler beam. For both the P\&M and StimPDC scheme, we measure the crosstalk matrices and calculate the QER for each basis. The results are averaged over 50 turbulence realizations, which is done by applying different phase masks. The corresponding data is presented in Fig.~\ref{fig:resultados}e, for $d=2$ and Fig.~\ref{fig:resultados}f, for $d=5$.

Our laboratory measurements are in strong agreement with our theoretical predictions. The discrepancies between numerical and experimental values can be attributed to measurement noise, with the signal-to-noise ratio estimated to be approximately 15 dB for $j=j'$. Even with these limitations, we demonstrate that fidelities can be improved by up to 30\% for the MUBs with $d=2$, and by up to 20\% for the MUBs with $d=5$ under the strongest turbulence conditions. More importantly, with our scheme, the average QER remains consistently below the corresponding $Q_{\text{max}}$ values for both MUBs across all turbulence levels studied. This result confirms the robustness of our approach in transmitting higher security key rates under adverse propagation conditions, highlighting its potential for practical implementations in quantum key distribution and high-dimensional quantum communication through turbulent channels.

\section{Conclusions}

We have proposed and demonstrated a novel quantum communication scheme based on StimPDC to mitigate the degrading effects of atmospheric turbulence in FSO high-dimensional QKD. By exploiting the intrinsic phase-conjugation properties of StimPDC, our scheme allows the spatial structure of a quantum signal to be self-corrected without requiring prior knowledge of the turbulent channel.

We developed a theoretical model to guide the optimal selection of spatial modes and system parameters for implementing the proposed scheme, and verified its performance through numerical simulations and a proof-of-principle experiment. Our results show that the proposed scheme significantly reduces the QER, which in turn increases the secure key rate of the protocol across different turbulence levels. In particular, we demonstrated improvements of up to 50\% over conventional P\&M approaches, maintaining QERs below the security threshold even under strong turbulence.

These findings highlight the potential of nonlinear optical techniques, particularly StimPDC, for robust and scalable quantum communication in realistic atmospheric conditions. Our work provides a pathway toward implementing high-dimensional QKD systems with improved noise resilience. Moreover, this technique could potentially be expanded to aberration correlation in microscopy and sensing applications based on information encoding with spatial modes of light, leading to novel schemes for quantum-enhanced imaging and metrology.

\section{Supplementary material}
\subsection{Finding the optimal basis}

Reference~\cite{aguilar} presents a detailed analysis describing how the fidelity of spatial modes generated via StimPDC depends on the parameters of the beams driving the process. The general idea is to identify a basis in which all components within the basis share the same optical beam diameter. A method to construct MUBs of OAM states is shown in Ref~\cite{mafu-2013}, where each basis element is an analytically defined, balanced superposition of positive and negative OAM states (Laguerre-Gaussian states with zero radial number).

To construct such bases, one must find the eigenbases of the Weyl operators $\{Z,XZ^l|l=0,1,...,d\}$, where\cite{mafu-2013}: 

\begin{equation}
    Z=\sum_{i=0}^{d-1}\omega^i \ket{i}\bra{i},
\end{equation}

\begin{equation}
    X=\sum_{i=0}^{d-1} \ket{i+\text{mod}d}\bra{i},
\end{equation}
here $\omega=e^{i2\pi/d}$. The eigenbasis of $Z$ corresponds to the standard OAM modes since it is diagonal in that basis ($\ket{i}=\{\ket{-l},...,\ket{-2},\ket{-1},\ket{0},\ket{1},\ket{2},...,\ket{l}\}$). However, these modes have different sizes and thus are less ideal for achieving uniform mode transfer fidelity via StimPDC. The only viable solutions are the eigenbases of the operators $\{XZ^l|l=0,1,...,d\}$.

Without loss of generality, we demonstrate our proposed scheme for $d = 2$ and $d=5$. As a proof of principle, we focus on the generalized BB84 protocol, which requires only two mutually unbiased bases. For $d = 2$, Weyl matrices reduce to the Pauli matrices. Then we select the eigenbases of $X$ and $XZ$, whose vectors correspond to the columns of the following matrices:

\begin{equation}
\label{MUB1d2}
\frac{1}{\sqrt{2}}
    \begin{pmatrix}
1 & 1\\
1 & -1 \\
\end{pmatrix},
\end{equation}

\begin{equation}
\label{MUB2d2}
\frac{1}{\sqrt{2}}
    \begin{pmatrix}
1 & -i\\
1 & i \\
\end{pmatrix}.
\end{equation}

For $d=5$, we select the eigenbases of $XZ$ and $XZ^3$, whose vectors are given by the columns of the following matrices:

\begin{equation}
\label{MUB1}
\frac{1}{\sqrt{5}}
    \begin{pmatrix}
\omega^{-2} & 1 & \omega & \omega & 1\\
\omega^{-1} & 1 & 1 & \omega^{-1} & \omega^2\\
1 & \omega^{-2} & 1 & \omega & \omega\\
\omega & \omega & 1 & \omega^{-2} & 1\\
\omega & 1 & \omega^{-2} & 1 & \omega\\
\end{pmatrix},
\end{equation}

\begin{equation}
\label{MUB2}
\frac{1}{\sqrt{5}}
    \begin{pmatrix}
1 & \omega^{-1} & \omega^{-1} & 1 & \omega^{2}\\
\omega^{2} & 1 & \omega^{-1} & \omega^{-1} & 1\\
1 & 1 & \omega & \omega^{-2} & \omega\\
\omega & \omega^{-2} & \omega & 1 & 1\\
1 & \omega & \omega^{-2} & \omega & 1\\
\end{pmatrix}.
\end{equation}

Here, Eq. \ref{MUB1} corresponds to MUB1, and Eq. \ref{MUB2} corresponds to MUB2. Finally, we compute the fidelity of each state, obtaining the results shown in Figure 2 in the main text.

\subsection{Turbulence phase mask for numerical and experimental test}
\begin{figure*}
\centering
    \includegraphics[width=1\linewidth]{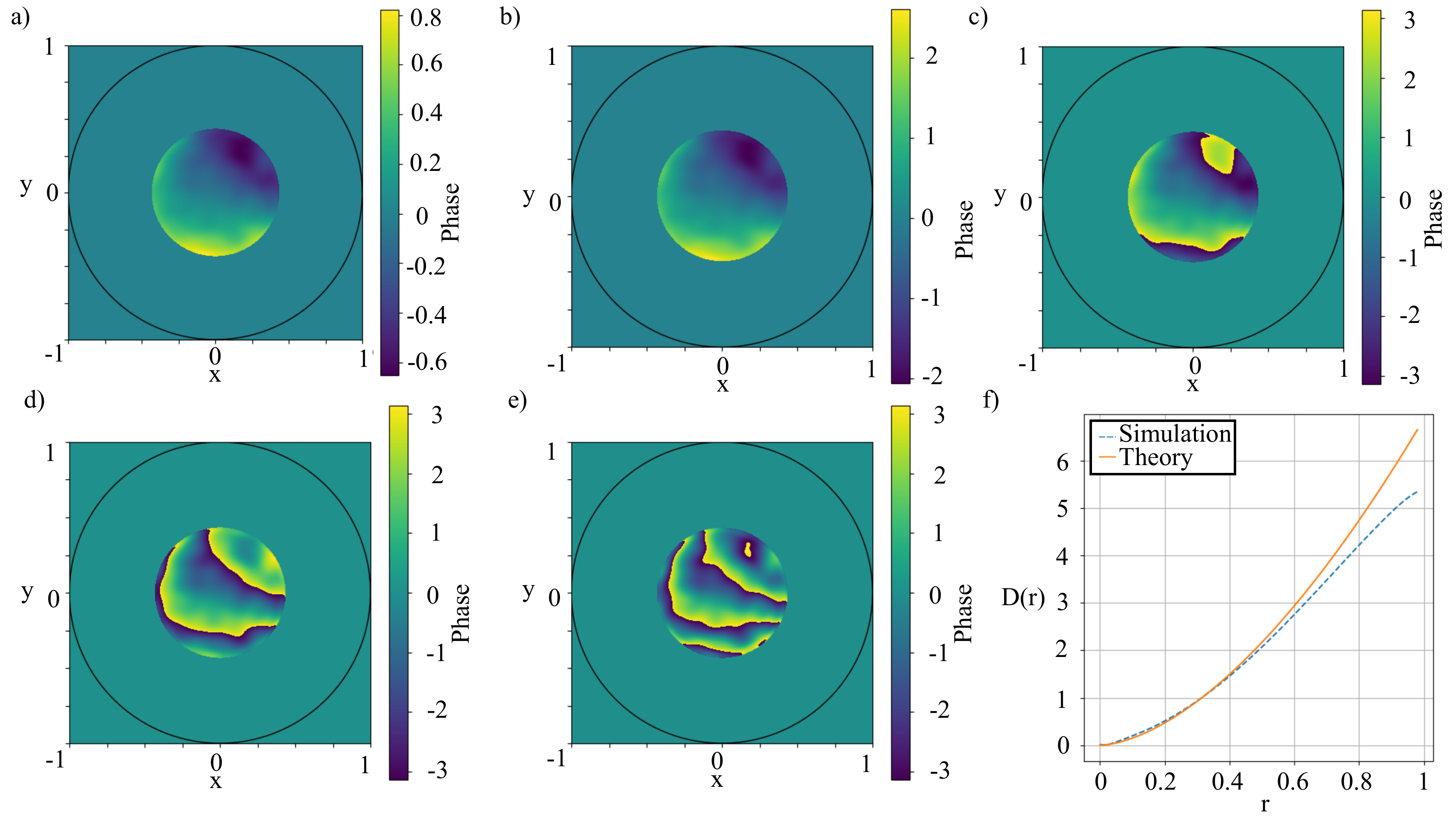}
    \caption{Examples of turbulent phase masks generated using Eq.\ref{zernike} with 172 terms. Panels a)–e) correspond to $D_B/r_0 =[1,5]$ , respectively. f) shows the structure function averaged over 100 generated masks using Eq.\ref{structure}. The results demonstrate that our numerical simulation agrees well with the theoretical function. }
    \label{fig:mascaras}
\end{figure*}
A Zernike polynomial representation of the transverse phase fluctuations induced by atmospheric turbulence can be expressed as~\cite{peters-2025}:

\begin{equation} 
\label{zernike}
\Theta(\rho, \phi) = \sum_j a_j Z_j(\rho, \phi).
\end{equation}
Here, the coefficients $a_j$ can be treated as Gaussian random variables with zero mean. Their statistical properties are characterized by the covariance matrix $\langle a_j a_{j'} \rangle$. This spectrum allows us to compute the covariance of the Zernike coefficients in the representation of the turbulent phase. Figure \ref{fig:mascaras}a-e shows examples of the phase masks generated by this method. 

The accuracy of phase screens generated by this or any other method is typically evaluated by how well they reproduce the expected phase structure function for a given turbulence model. The structure function of the phase fluctuations is defined as:

\begin{equation}
\label{structure}
D(r) = \langle \left( \Theta(\mathbf{r}) - \Theta(\mathbf{r} - \mathbf{r}_1) \right)^2 \rangle = 6.88 \left( \frac{r}{r_0} \right)^{5/3}, 
\end{equation}
here, $r_0$ is the Fried parameter, defined in the main text. Figure \ref{fig:mascaras}f shows the structure function of an average of 50 our generated phase mask.








The numerical method to simulate the propagation of the beams in turbulent media is based on an iterative process that alternates between free-space propagation (given by the spectrum method) of the beam and the application of a turbulent phase mask, as generated by the method described previously. This process is repeated over successive segments along the total propagation path. The full propagation distance is divided into shorter segments such that the partial Rytov parameter for each segment satisfies the condition $\sigma_{R}^2=1.23C_n^2 k^{7/6} Z_T^{11/6}<1$~\cite{peters-2025}. This criterion ensures the validity of the Rytov approximation, which assumes small phase modulations of the initial beam. Statistical reliability is achieved by averaging all observables over a large number of independent turbulence realizations. The results for each data point displayed in the main text are obtained by averaging over the results of 50 simulations. 

As an example, in our simulations with parameters $D_B/r_0=$ 4, where $D_B=$ 6 cm, $Z_T=$ 1 km, $\lambda=810$ nm and $C_n^2=4.2\times10^{-14}$ m$^{2/3}$. The total Rytov parameter is $\sigma_{R,T}^2=1.78$. To stay within the validity range of the Rytov approximation, the propagation path is divided into four segments (therefore we use 4 different turbulent phase mask for each simulation), each with a partial Rytov parameter of $\sigma_{R,P}^2=0.14<1$. The screen size is 0.24 m with a grid spacing of $\Delta x=\Delta y$ = 0.0002 m, and we adjust the diameter of the phase screens to the optical diameter of the beams. 

Following the same criterion in our experiment, where only a single phase mask is available, we ensure that the beam size $D$ is chosen such that the experiment can be performed within the same $D_B/r_0$ range used in the simulations. For the maximum turbulence strength in our experiment, $D_B/r_0 = 5$, we have $D_B = 0.5$ cm, $Z_T = 1$ m, $\lambda_i = 840$ nm, and $C_n^2 = 4.2 \times 10^{-9}$, which yields a total Rytov parameter of $\sigma_{R,T}^2 = 0.54<1$. This confirms the validity of our experimental results.

\subsection{Measuring QER}
\begin{figure*}
\centering
    \includegraphics[width=0.8\linewidth]{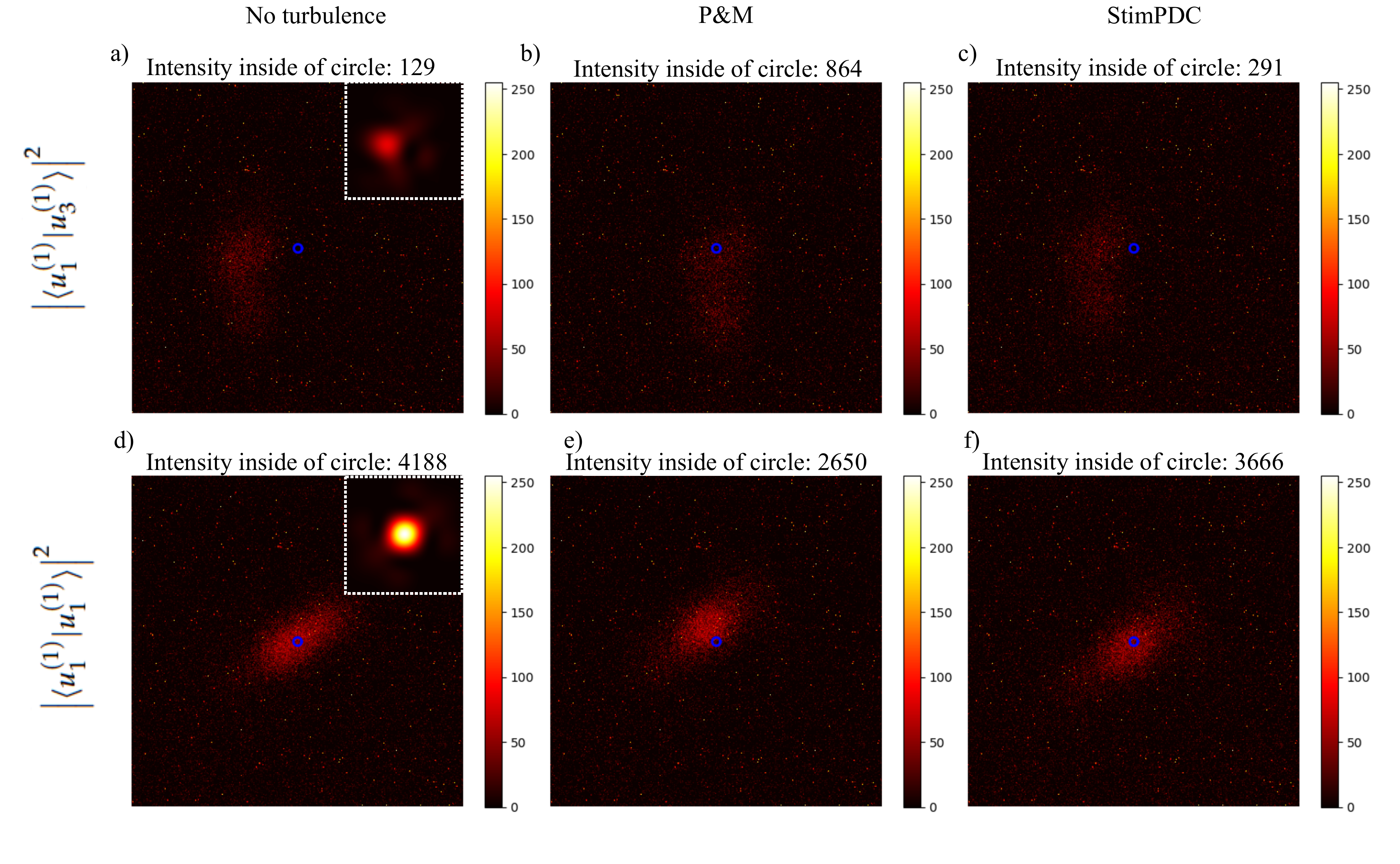}
    \caption{Examples of output intensities at the CMOS plane. The circle indicates the center of the optical axis, within which the intensity is integrated to approximate $\left| \braket{u_{Bob}|u_{Alice}} \right|^2$. The first row shows $\left| \braket{u_1^{(1)}|u_{3}^{(1)}} \right|^2$, while the second row shows $\left| \braket{u_1^{(1)}|u_{1}^{(1)}} \right|^2$, both for $d=5$. Panels (a) and (d) correspond to measurements without turbulence, with insets displaying the Fourier-plane intensity. Panels (b) and (e) show the P\&M method, and panels (c) and (f) correspond to the StimPDC method with $D_B/r_0 = 3$. All values are normalized to the original fidelity (first row), allowing the extraction of both the mode fidelity and the elements of the crosstalk matrix.}
    \label{fig:QER1}
\end{figure*}

To experimentally measure the overlap $\left| \braket{u_j^{(b)}|u_{j'}^{(b')}} \right|^2$, we follow the technique reported in Ref.~\cite{pinnell-2020}, which employs the Fourier transform of the mode transmitted by Alice, projected onto SLM2. 

First, as shown in the main text, we generate the spatial states using StimPDC. Then, with the aid of the telescope formed by lenses $f_5$ and $f_6$, we project the beam onto SLM2, where Alice encodes the spatial mode she wants to measure. The resulting field at this plane is given by $U_i(\vec{r}) U_{\text{SLM2}}^*(\vec{r})$. Next, using the Fourier lens $f_7$, we obtain the Fourier transform of the field as $\int U_i(\vec{r}) U_{SLM2}^*(\vec{r}) \text{exp}\{-\frac{i2\pi}{\lambda_i f_7}\vec{q} \cdot \vec{r}\} d\vec{r}$. As examples, in the insets of Fig.~\ref{fig:QER1}a and \ref{fig:QER1}d, we show the Fourier field for dimension $d=5$, MUB1, $j=1$, with projection in $j'=3$ and $j'=1$, respectively. Now, the field intensity on the optical axis ($\vec{q}=0$) is $\int U_i(\vec{r}) U_{SLM2}^*(\vec{r}) d\vec{r}\propto\braket{u_j^{(b)}|u_{j'}^{(b')}} $, which enables us to measure the fidelity and crosstalk matrices. As an example, in the case of $\left| \braket{u_1^{(1)}|u_{3}^{(1)}} \right|^2$, shown in the inset of Fig.~\ref{fig:QER1}a, the theoretical value at the optical center is zero, since the states are orthogonal. For the case $\left| \braket{u_1^{(1)}|u_{1}^{(1)}} \right|^2$, shown in the inset of Fig.~\ref{fig:QER1}d, the theoretical optical center value is one.

In our experiment we take measurements of the field intensity with a CMOS camera (pixel size 3.45 $\mu$m $\times$ 3.45 $\mu$m). In the case of $\left| \braket{u_1^{(1)}|u_{3}^{(1)}} \right|^2$, shown in Fig.~\ref{fig:QER1}a, we define a small integration region at the center of the optical axis, that captures the average background noise for each state, with its size determined by the Rayleigh criterion~\cite{pinnell-2020}. The circle employed has a diameter of 5 pixels (17 $\mu$m). The integrated value within this circle corresponds only to background noise, which we then subtract from the final results. For the case $\left| \braket{u_1^{(1)}|u_{1}^{(1)}} \right|^2$ shown in Fig.~\ref{fig:QER1}d, the integrated intensity within the circle corresponds to the value used to normalize all our fidelity measurements, since this is the maximum recorded intensity. We repeat this procedure for all states, which allows us to construct both the fidelity and correlation matrices, as defined in the main text. The results are presented in Fig.~6c of the main manuscript.

Using the same measurement procedure, we can obtain the fidelity and crosstalk matrices for the cases with turbulence. In the P\&M scheme, the turbulence masks described in the previous section are applied only to the stimulated idler beam, while no phase distortion is applied to the seed beam. Figures~\ref{fig:QER1}b and \ref{fig:QER1}e show examples of $\left| \braket{u_1^{(1)}|u_{3}^{(1)}} \right|^2$ and $\left| \braket{u_1^{(1)}|u_{1}^{(1)}} \right|^2$, respectively, under turbulence strength $D_B/r_0 = 3$. We observe that, under the P\&M method, the beam shape is distorted and the intensity centroid is displaced. As a result, for non-orthogonal modes such as in Fig.~\ref{fig:QER1}b, the on-axis intensity increases, leading to crosstalk between modes. Conversely, when the transmitted and projected modes are the same, as in Fig.~\ref{fig:QER1}e, the on-axis intensity decreases, thereby reducing the mode fidelity.

Finally, when the turbulence masks are also applied to the seed beam, the spatial phase structure is conjugated onto the stimulated beam, thereby correcting the fidelity and crosstalk matrices. Figures~\ref{fig:QER1}c and \ref{fig:QER1}f show examples of $\left| \braket{u_1^{(1)}|u_{3}^{(1)}} \right|^2$ and $\left| \braket{u_1^{(1)}|u_{1}^{(1)}} \right|^2$, respectively, under turbulence strength $D_B/r_0 = 3$ using our proposed method. As we can see, for non-orthogonal modes such as in Fig.~\ref{fig:QER1}c, the on-axis intensity decreases compared to the P\&M case, leading to reduced crosstalk between modes. Moreover, when the transmitted and projected modes are the same, as in Fig.~\ref{fig:QER1}f, the on-axis intensity increases, thereby improving mode fidelity. It is worth noting that, although these corrections are effective, the values are not fully identical to those obtained in the absence of turbulence, as discussed in the main text.

This procedure is carried out for each mode of every basis in each dimension, allowing us to construct the fidelity and crosstalk matrix for each basis. From these matrices we extract the loss of fidelity $\Delta F$ and QER. The procedure is repeated 50 times to ensure sufficient statistical sampling, yielding the results shown in Figs.~5c and 5d of the main text.

\bibliography{sample}


\end{document}